\documentclass[review]{elsarticle}
\usepackage{lineno,hyperref}\usepackage[dvipsnames]{xcolor}
\usepackage{ulem}
\usepackage{subcaption}
\usepackage{xcolor}
\usepackage[dvipsnames]{xcolor}
\usepackage{soul}

\definecolor{Seashell}{RGB}{255, 245, 238} %background lighten
\definecolor{Firebrick4}{RGB}{255, 0, 0}% red text

\usepackage{graphicx}
\usepackage{multirow}
\usepackage{booktabs}
\usepackage[linesnumbered,ruled,vlined]{algorithm2e}
\usepackage{mathtools}

\usepackage{comment}

\journal{Journal of \LaTeX\ Templates}

\begin{document}
	
	\begin{frontmatter}
		
		\title{A Few-Shot Meta-Learning based Siamese Neural Network using Entropy Features for Ransomware Classification}
		
		
		\author[mymainaddress]{Jinting Zhu \corref{mycorrespondingauthor}}
		\ead{jzhu3@massey.ac.nz}
		
		\author[mymainaddress]{Julian Jang-Jaccard}
		\author[mymainaddress]{Amardeep Singh}
		\author[secondaddress]{Ian Welch}
		\author[secondaddress]{Harith AI-Sahaf}
		\author[thridaddress]{Seyit Camtepe}
		
		\cortext[mycorrespondingauthor]{Corresponding author}

		\address[mymainaddress]{Cybersecurity Lab, Massey University, Auckland, NEW ZEALAND}
		\address[secondaddress]{School of Engineering and Computer Science, Victoria University of Wellington, Wellington, NEW ZEALAND}
		\address[thridaddress]{Data61, CSIRO, AUSTRALIA}

	%%%%%%%%%%%%%%%%%%%%%%%%%%%%%%%%%%%%%%%%%%%%%%%%%%%%%%%%%%%%%%%
	%%
	%%  Abstract
	%%
	%%%%%%%%%%%%%%%%%%%%%%%%%%%%%%%%%%%%%%%%%%%%%%%%%%%%%%%%%%%%%%%	
		\begin{abstract}
			Ransomware defense solutions that can quickly detect and classify different ransomware classes to formulate rapid response plans have been in high demand in recent years. Though the applicability of adopting deep learning techniques to provide automation and self-learning provision has been proven in many application domains, the lack of data available for ransomware (and other malware) samples has been raised as a barrier to developing effective deep learning-based solutions. To address this concern, we propose a few-shot meta-learning based Siamese Neural Network that not only detects ransomware attacks but is able to classify them into different classes. Our proposed model utilizes the entropy feature directly obtained from ransomware binary files to retain more fine-grained features associated with different ransomware signatures. These entropy features are used further to train and optimize our model using a pre-trained network (e.g. VGG-16) in a meta-learning fashion. This approach generates more accurate weight factors, compared to feature images are used, to avoid the bias typically associated with a model trained with a limited number of training samples. Our experimental results show that our proposed model is highly effective in providing a weighted F1-score exceeding the rate \textgreater 86\% compared to other similar methods.

 %This shows that our proposed approach can be used as a ransomware triage solution that can quickly screen ransomware attacks to address the serious concern raised for the mass growth of ransomware attacks worldwide and their disastrous consequences.
		\end{abstract}
		
		
	\end{frontmatter}

	%%%%%%%%%%%%%%%%%%%%%%%%%%%%%%%%%%%%%%%%%%%%%%%%%%%%%%%%%%%%%%%
	%%
	%%  Introduction
	%%
	%%%%%%%%%%%%%%%%%%%%%%%%%%%%%%%%%%%%%%%%%%%%%%%%%%%%%%%%%%%%%%%
	\section{Introduction}
	\label{sec:introduction}
	
Ransomware is a type of malware designed to attack the victim's system and denies access to the victim’s sensitive data until a ransom is paid \cite{DAVIES2021102377}.  With the popularity of bitcoin that hides the true identity of the account used by attackers along with the prevalence of encryption techniques, this type of malware has proliferated recently, causing millions of dollars in losses for businesses and consumers.

There are two approaches for performing ransomware (and other malware) detection: static and dynamic-based analysis. The dynamic analysis-based detection method can provide a high detection rate \cite{Alsahaf19, Abbasi2020}, but requires the execution of malicious code.  However, this may cause a serious scalability issue to process the large number of previously unseen binaries that come across the desk of malware analysts. In addition, they are not able to extract significant behavioral patterns of the ransomware due to anti-emulation techniques such as fingerprinting or time bombs \cite{RHODE2018578}. Furthermore, different ransomware families vary significantly thus make difficult to profile their behavior pattern \cite{DAVIES2021102377, camp2019measuring}. Static analysis overcomes some of the dynamic analysis detection issues by detecting ransomware attacks prior to their execution. This is done by analyzing bytes or instruction sequences to carry out the malicious activity (i.e., the signature of ransomware in the code).  For this reason, we utilize static code analysis to obtain unique features of ransomware despite the obvious advantages present in the dynamic analysis.

With the popularity of Artificial Intelligence (AI) techniques, many (shallow) machine learning and deep learning methods have been proposed. We argue that there are two significant drawbacks to the existing state-of-the-art.
The first issue is the majority of existing research tends to focus on learning from the generic features of malware images from different malware families and building feature profiles \cite{Nataraj2011, kim2020,lichen, 8763852, Yue08042}. Here the feature profiles represent the (static) features contained in a malware sample to provide an important clue as to whether the image sample is malicious or not. %\textcolor{BurntOrange}{The malware image conversion technique proposed by \cite{Nataraj2011} are commonly used in malware classification task. However, there are some  disadvantages are existed in the malware image. Firstly, the number of white noises \cite{catak2021data} are generated from the android executable files, which result in the classification ability of deep learning model decreased. Secondly, the malware images as input have to downsampling to same shape so that the information of original files will be loss. Thirdly, it has higher computational \cite{han2015malware} compared with entropy feature based.}

We also argue that there are issues with using malware grayscale images as feature profiles. For example, the obfuscation and repackaging used to introduce new ransomware variants is treated as a form of noise when creating the grayscale image. This often makes it impossible to accurately classify many variants that come from the same malware family  \cite{milosevic2017machine,vasan2020image,yuan2020byte}. Secondly, malware grayscale images of executables are of different sizes and shapes. In order to use them as input to the deep learning model, they need to be downsampled to the same shape, resulting in a loss of information compared to the original file. Finally, the computational cost \cite{han2015malware} of training a model based on grayscale images is higher compared to entropy-based features.
Moreover, only using the common static information (e.g., program byte sequences, instructions in the form of disassembled opcodes, functional calls) as data input points to feed into a deep neural network cannot capture unique characteristics of each malware family signature. This shortcoming limits the use of static information not able to accurately classify many variants driven from the same malware family \cite{milosevic2017machine, vasan2020image, yuan2020byte}.

 The second issue with the existing approaches is the demand for large data inputs to find more relevant correlations across the features \cite{zhu2021joint}. They are unable to detect and classify the malware families trained with a limited number of samples \cite{cao2018softmax, zhu2020multi, zhu2021task}.

In this research work, we propose a few-shot learning based method Siamese Neural Network (SNN) that is capable of not only detecting different ransomware families but also accurately classifying ransomware variants belonging to the same ransomware family even in the presence of only a small number of training samples available.

	The contributions of our proposed model are the following:
	\begin{itemize}
		\item Our proposed model can learn unique ransomware signatures from different ransomware classes even if the number of ransomware samples for each class is very small.
\item Different from the existing models, which typically use image features, we use entropy features directly obtained from each ransomware binary file as inputs to our model. This prevents information loss typically associated with image feature conversion. The use of entropy features thus significantly contributes towards accurately distinguishing different ransomware signatures represented in different ransomware classes, addressing the first issue we discussed earlier.
		\item To solve the second issue associated with the existing approaches, we use a pre-trained VGG-16 network as a part of the meta-learning process to generate weights that more accurately capture the characteristics of each ransomware sample. This not only contributes to improving the classification accuracy but also avoids the potential bias associated with the use of a limited number of training samples in a deep learning model.
\item Our proposed model uses a combination of two center losses and a softmax loss to accurately capture the similarity between the ransomware samples belonging to the same class (i.e., intra-class variance) and the di-similarity across the ransomware samples across different ransomware classes (i.e., inter-class variance).
		\item Our experiments, tested on a total of 1,046 ransomware samples with 11 different classes, show that the proposed model is highly effective for correctly classifying different ransomware samples by achieving the classification with the weighted F1-score exceeding 86\%, which outperformed other existing benchmark methods.
	\end{itemize}

	The rest of the paper is organized as follows. Section \ref{sec:related_work} presents related works for ransomware detection and classification. Section \ref{sec:materials_and_methodlogy} provides the details of the proposed  model. Section \ref{sec:experimental_setup_and_evaluation_metrics}, describes the experimental setup and evaluation metrics. In Section \ref{sec:results_and_discussion}, the experimental results are presented and discussed. Lastly, Section \ref{sec:conclusions} draws a conclusion from our proposed method and describe the future work that is planned.

	%%%%%%%%%%%%%%%%%%%%%%%%%%%%%%%%%%%%%%%%%%%%%%%%%%%%%%%%%%%%%%%
	%%
	%%  2. Related Work
	%%
	%%%%%%%%%%%%%%%%%%%%%%%%%%%%%%%%%%%%%%%%%%%%%%%%%%%%%%%%%%%%%%%
	\section{Related work}
	\label{sec:related_work}
	
	Dynamic analysis avoids the problem of having to preprocess samples using an unpacker. However, it has several limitations. First, there it requires either partial or full execution of the malware itself using an emulator or sandbox. Second, the malware sample may use anti-virtualization techniques that detect and prevent its execution in a sandbox or have particular conditions before running that are not triggered in a sandbox environment. Third, there is no guarantee that you will achieve sufficient code coverage for the purposes of accurate classification.
	
	Static analysis based upon reverse engineering also suffers from costs associated with dealing with obfuscation or performing data or control flow techniques. Therefore, researchers have looked for effective but efficient static analysis techniques that do not require reverse engineering such as disassembly of the malware samples. Nataraj et al. 2011 \cite{Nataraj2011} proposed a less expensive but effective static analysis technique based upon image processing methods. Their central insight was to treat binary programs like images and use a visual similarity to avoid either execution or disassembly of the malware. Their preliminary work showed promising results, and other researchers have subsequently expanded upon their ideas. 	More recently Kim et al. \cite{kim2020} proposed a binary classification system that takes a hybrid approach that extracts features from the portable executable (PE) file as well as using both black and white and color images created from the file's binary representation. The features derived from the PE file header and sections include entropy and packers, where the packers are classified using YARA rules. Decision tree, random forest, and gradient boosting algorithms are used to classify the malware based upon the features from the PE file.
	
	Several machine learning approaches have been attempted to automate the learning of different malware signatures. SigMal \cite{Kirat13} extracts features from the PE structure of malware executable. SigMal transforms these features into a digital image. Signatures are extracted from images and investigated with similarity measures using the KNN technique. They claimed above 99\% precision evaluated on the samples they collected. Baldwin and Dehghantanha \cite{Baldwin2018} used static analysis to extract opcode characteristics as input features for a support vector machine (SVM) classifier to classify five class ransomware families with $96.5\%$ accuracy. In the same vein, Zhang et al. \cite{ZHANG2019211} transformed raw opcode sequences into N-gram-based input features for a random forest algorithm to achieve $91.43\%$ accuracy. The downside of these techniques is they need to use reverse engineering with two main problems - scalability to convert binary representations into opcodes and anti-disassembly techniques used to generate incorrect opcode sequences.
	
	To capitalize image-based features proposed by \cite{Nataraj2011}, various end-to-end deep learning solutions based on convolutional neural network (CNN) architecture are proposed are in literature \cite{8330042,CUI201950}. Furthermore, these works were extended to have a capability of CNN under limited input samples available for training with transfer learning such as Chen \cite{lichen} who proposed the training of their model with Inception neural network's pre-trained weights. In this work,  low-level features were borrowed from the Inception model and applied that transferred knowledge to malware classification. In the same vein, Lo et al. \cite{8763852} employed pre-trained weights from Xception neural network with malware classification. Likewise, Yue \cite{Yue08042} proposed weighted softmax loss for the CNN model to classify malware in an unbalanced dataset along with VGG pre-trained weights to fine-tune the CNN model.

	The static analysis offers early detection of ransomware but sometimes ransomware use obfuscation and polymorphism techniques that avoid signature detection through static analysis \cite{mcintosh2018large}. To address this, various hybrid solution is proposed that combines static and dynamic techniques such as system resource usages and statistics with opcode frequency \cite{Ferrante,8328219}. A CNN algorithm was used to classify the malware based upon the images. A final decision algorithm uses majority voting to come to a final classification as malicious or benign. This approach achieved promising accuracy but requires an extra step of preprocessing the PE file itself whereas we focus on a single image and can deal with the scarcity of training data.   Zhu et al. \cite{zhu2020multi} proposed an SNN-based model to classify different android malware types. They use a batch normalization layer and multiple loss functions to address the overfitting issue associated with a model trained with a limited number of samples. These existing models tend to be effective at capturing malware samples when the common attribute of malicious code is distinct but often do not consider when there are slight differences in features of the malware as is the case of many obfuscated malware variants.

%%%%%%%%%%%%%%%%%%%%%%%%%%%%%%%%%%%%%%%%%%%%%%%%%%%%%%%%%%%%%%%
	%%
	%%  3. Materials and Methodology
	%%
	%%%%%%%%%%%%%%%%%%%%%%%%%%%%%%%%%%%%%%%%%%%%%%%%%%%%%%%%%%%%%%%
	\section{Materials and methodology}
	\label{sec:materials_and_methodlogy}
	
\begin{figure}[h]
\centering
\includegraphics[scale=1.0]{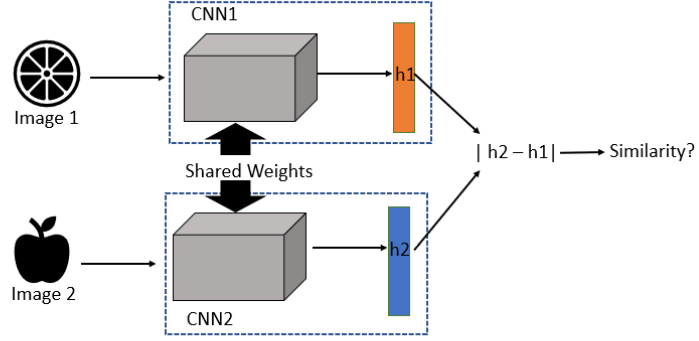}
\caption{The overview of Siamese Neural Network}
\label{fig:snngeneric}
\end{figure}

Siamese Neural Network (SNN) has been applied in many applications such as computer vision and natural language processing. Its core function involves estimating the similarity between two images represented by feature embedding space and generating a similarity score. A generic SNN is comprised of two sub-networks, as shown in Figure \ref{fig:snngeneric}, where the weights and hyperparameter settings are shared. Both sub-networks take one image belonging to the same class (i.e., positive pairs) as input and output the features it learned. These features are projected in the feature embedding of the fully connected layer while a loss function is used to predict whether the images processed by two sub-networks belong to the same class or not.
It can also predict if the different features learned are complementary to different instances within the same class.
%because there are slight differences between each sample of the same class.  Through the combination of features that can make up the feature lack of spatial distribution information in the entropy graph and reduce the impact of inherent defects of a single feature.}
For example, it can correctly predict that different ransomware instances still belong to the same ransomware family as they share the unique ransomware signature.
   %Through the combination of features that can make up the feature lack of spatial distribution information in the entropy graph and reduce the impact of inherent defects of a single feature.

	%% +++++++++++++++++++++++++++++++++++++++++++++++++++++++
	%%
	%% Entropy Features
	%%
	%% +++++++++++++++++++++++++++++++++++++++++++++++++++++++
	\subsection{Entropy Feature}
	\label{ssec:entropy_features}
	
	The main innovation of our proposed model is using entropy features with SNN to train and optimize the model in a meta-learning fashion that utilizes already trained machine learning algorithms (e.g., VGG-16). The entropy features we use are directly computed from binary ransomware files therefore they retain unique characteristics involved in each ransomware signature much better, compared to image features \cite{Nataraj2011} typically used in the existing state-of-the-art. This has a clear advantage over using image features as inputs which tends to lose precise information through conversion processes in order to produce the size of images that works better for a specific model. By using the entropy feature, any computations our model goes through to find feature correlations across different ransomware samples become more robust compared to when the image features were used. This is because image features can be easily influenced by white noise and complex texture features while entropy features are less sensitive to the spatial distribution of the small changes made to the generic code thus able to recognize any arbitrary changes better (i.e., obfuscation methods applied to avoid detection) \cite{hamid2018evaluation, zhu2020multi, zhu2021task}. Thus, the entropy feature resulted in more accurate classification and is especially useful to find ransomware variants derived from the generic code of the same ransomware family.

\begin{figure}[th!]
    	\centering
    	\subfloat[dalexis class] {\includegraphics[width=0.45\textwidth]{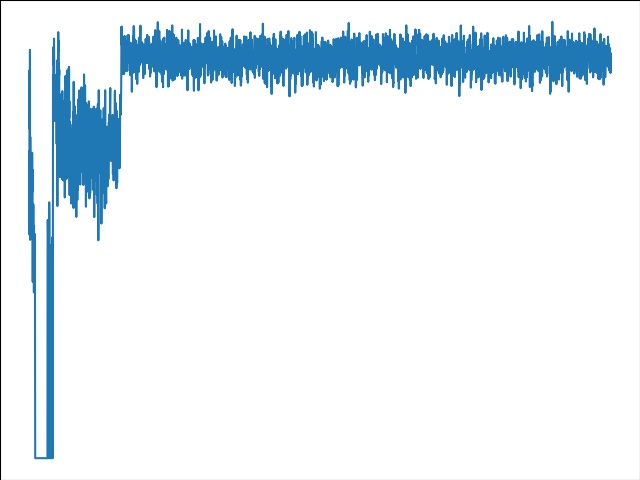}} \hfil
    	\subfloat[dalexis class] {\includegraphics[width=0.45\textwidth]{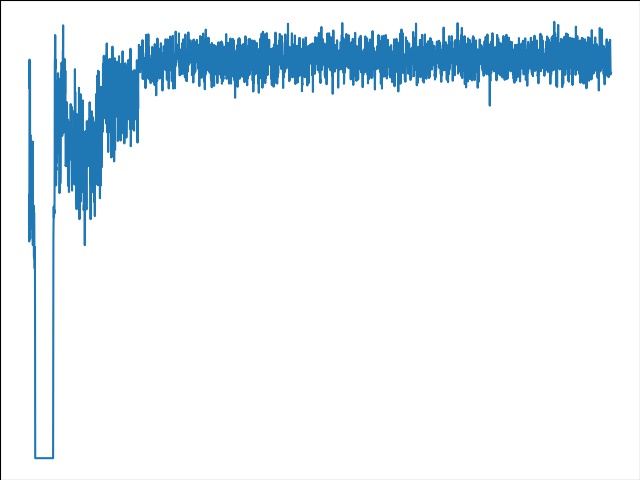}} \hfil
    	\subfloat[WannaCry class] {\includegraphics[width=0.45\textwidth]{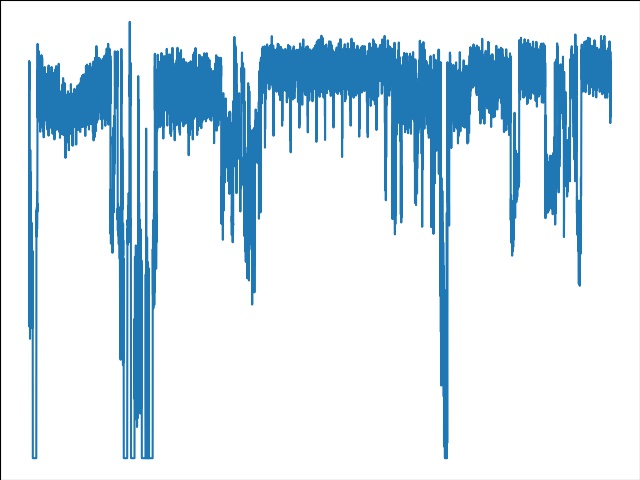}} \hfil
    	\subfloat[WannaCry class] {\includegraphics[width=0.45\textwidth]{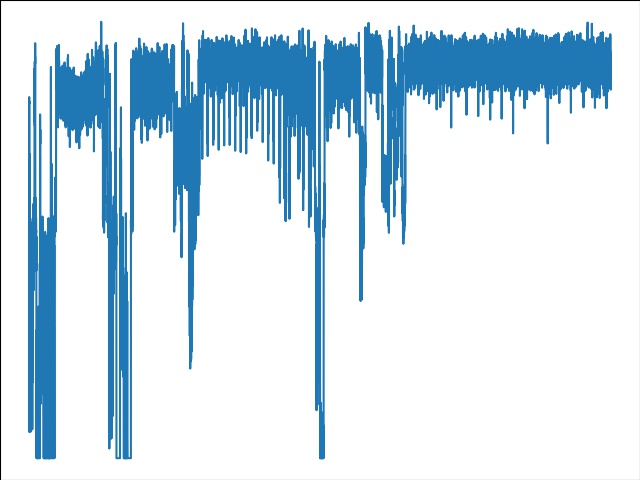}}
    	\caption{Examples of entropy graphs of two ransomware families.}
    	\label{fig:entropy}
\end{figure}

To illustrate the usefulness of using entropy features, Figure \ref{fig:entropy} shows the entropy graphs, containing entropy values from entropy features, obtained from two different ransomware families each of which has two variants. Figures \ref{fig:entropy}(a) and \ref{fig:entropy}(b) are two ransomware variants from the same ransomware family “dalexia” while Figures \ref{fig:entropy}(c) and \ref{fig:entropy}(d) are two ransomware variants from the ransomware family “WannaCry”. There is a significant difference in the shape of the entropy graphs between two families thus providing very unique feature information when trained. Among the variants of the same ransomware family, there is a slight difference in the entropy graphs but the general pattern shown in each entropy graph is similar to each other thus our model recognizes them as the instances belonging to the same ransomware family.

	\begin{figure}[th!]
    	\centering
    	\subfloat[] {\includegraphics[width=0.49\textwidth]{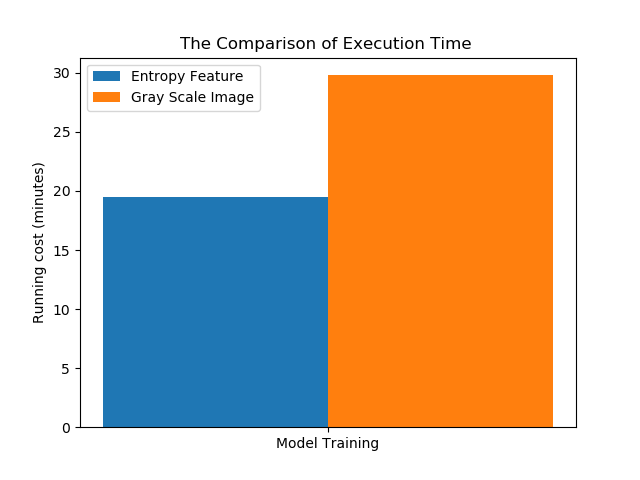}} \hfil
    	\subfloat[] {\includegraphics[width=0.49\textwidth]{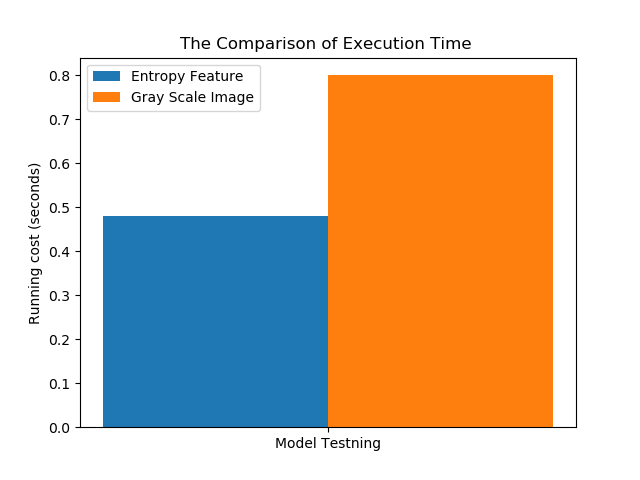}} \hfil
    	\caption{The comparison of execution time with or without entropy feature.}
    	\label{fig:execution_time}
    \end{figure}

\textcolor{blue}{We tested the computational cost by comparing the training time of our proposed model with entropy features and with grayscale images. As shown in Figure \ref{fig:execution_time}, (a) the proposed model based on entropy features takes about 10 minutes less time to train than the model based on grayscale images. (b) The execution time of the model based on entropy features for a new test example is less than 2 seconds compared to the model based on grayscale images.}

Our model utilizes the entropy values extracted from the entropy graph to train our proposed model to find different ransomware classes. The overall algorithm involved to produce an entropy graph from a ransomware binary file is depicted in Algorithm \ref{alg:entropy_graph}.
\begin{algorithm} [th!]
		\SetKwInOut{Input}{Input}
		\SetKwInOut{Output}{Output}
		
		\Input{$f$: malware binary file; $l$: segment length; $n$: the number of files }
		\Output{entropy graph matrix $m$}
		
		\While{not reach to n}{
			1. read $l$ bytes from $f$, and defined as segment $s$;
			
			2. \textbf{for} $i = 0$ to $255$ \textbf{do}
			
			3. \ \ \  compute the probability $p_i$ of $i$ appearing in $s$;\\
			
			4. \ \ \  compute the Shannon entropy
		}
		
		Generate entropy graph $m$\ \\
		%3) The entropy graph is fed to a pre-trained network (e.g., VGG-16) and  a feature vector is extracted from the image in Step 2; \\
		\caption{Generating Entropy Graph}
		\label{alg:entropy_graph}
	\end{algorithm}

To construct an entropy graph, we first read a ransomware binary file as a stream of bytes. The bytes are made into multiple segments each of which is comprised of 200 bytes. We further count the frequency of unique byte value followed by computing the entropy using Shannon’s formula as follows:
	\begin{equation}
		\mathit{Ent} = -\sum_{i}\sum_{j}M(i,j)logM(i,j)
	\end{equation}\label{eq:entropy}
	where $M$ is the probability of an occurrence of a byte value.

The entropy obtains the minimum value of 0 when all the byte values in a binary file of ransomware are the same while the maximum entropy value of byte value 8 is obtained when all the byte values are different.  The entropy values are then concatenated as a stream of values which can be formed as an entropy graph.

	%% +++++++++++++++++++++++++++++++++++++++++++++++++++++++
	%%
	%% 3.2 Processed SNN Structure
	%%
	%% +++++++++++++++++++++++++++++++++++++++++++++++++++++++
	\subsection{Our proposed Model}
	\label{ssec:processed_snn_structure}
	
\begin{figure}[th!]
		\centering
		\includegraphics[width=1.0\textwidth]{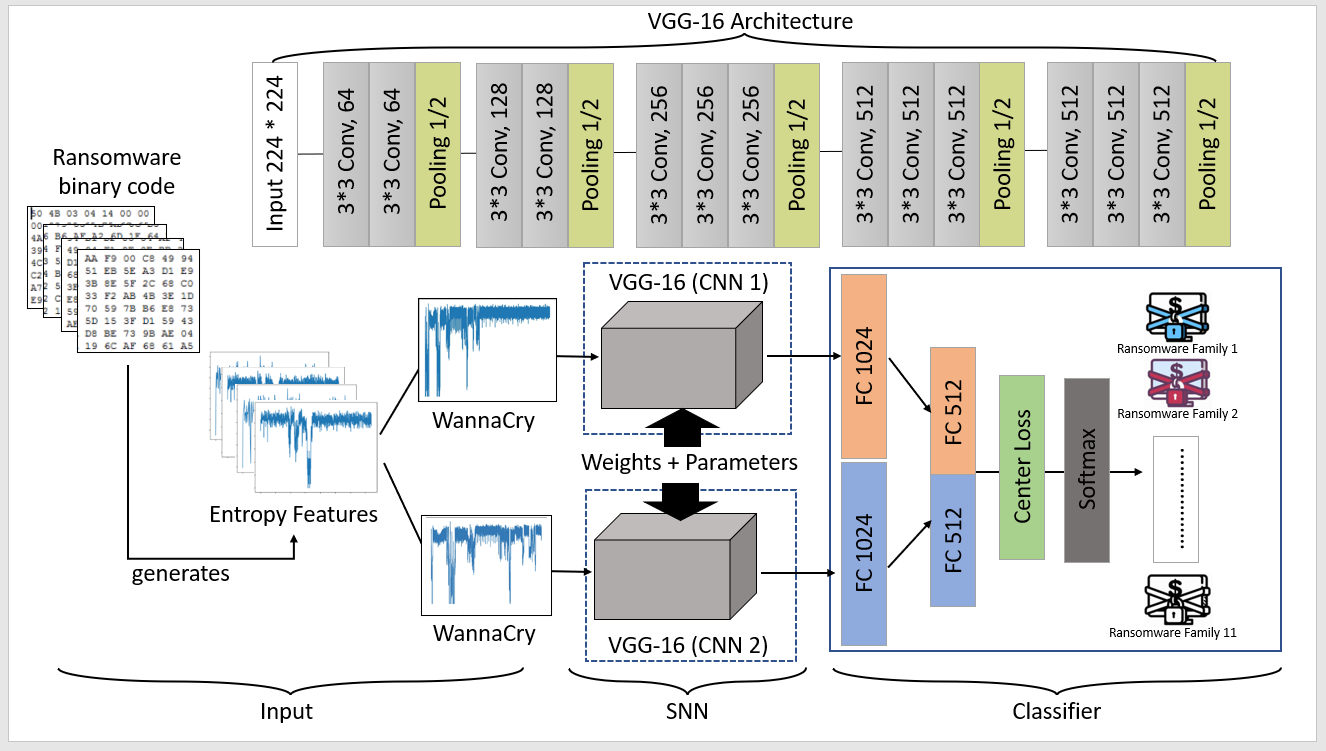}
		\caption{The overview of our proposed SNN using Entropy features}
		\label{fig:overview}
	\end{figure}

The overview of our proposed model is shown in Figure \ref{fig:overview}. The entropy values from each entropy graph are fed into each sub-network of SNN. At each sub-network, we use a pre-trained VGG-16  whose weights and parameters were trained on ImageNet and use it in a meta-learning fashion (i.e., the pre-trained model assists the training of our proposed model).

The VGG-16 architecture we use has 5 blocks each of which comprises several convolution layers and a pooling layer. The first two blocks contain 2 convolution layers of the receptive size of 3 $\times$ 3 with ReLU activation functions. The first block contains the convolution layer with 64 filters while the second block contains the convolution layer with 128 filters. The next three blocks contain 3 convolution layers, again with the receptive size of 3 $\times$ 3 with ReLU activation functions. The convolution layer in the third block contains 256 filters while the last two blocks (i.e., fourth and fifth) contain 512 filters at each convolution layer. The convolution stride is fixed at 1 pixel as well as the padding size at 1 pixel. We use max-pooling at each block over a 2 $\times$ 2 pixel window with a stride of 2 pixels.

This pre-trained VGG-16 architecture is further trained and optimized with the entropy values inputted as 2-dimensional vectors of a fixed size 224 $\times$ 224. By first utilizing weights and parameters well trained with the image samples and further fine-tuning the parameters of our proposed model with entropy features. This way can avoid potential bias associated with training a deep learning model with a limited number of samples. Note that each sub-network takes one entropy graph as input from the same ransomware family.

The five blocks of VGG-16 architecture are followed by two fully connected layers with a fattening layer in-between. The first fully connected layer has 1024 neurons and the last fully connected layer serves as the output layer and has 512 neurons. We apply a center loss function to the final output layer to compute a loss across the inter-class variance (e.g., the combination features with two sub-networks). Finally, a softmax loss is used for categorical classification to distinguish different ransomware families.

Algorithm \ref{alg:alg1} illustrates the pseudocode involved in our proposed model in terms of training and testing stages.
	
	\begin{algorithm}[hbt!]
		\SetKwInOut{Input}{Input}
		\SetKwInOut{Output}{Output}
		
		\Input{Entropy Graph Feature $x_i$, Sample label $y_t^i$, Hyper-parameters $\beta$ , Initialized Centers $c_{y_t}$ and Class Weights $w_{y_t}$}
		\Output{Class Probability of Each Class}
		
		Dataloader = SelectPostivePairs($y_t^i$)\\
		
	   for $x_1$, $x_2$ in Dataloader: \\
	   \ \ \ \ \     $z_t^1,z_t^2$ = EncoderNetworks$(x_t^1,x_t^2)$ \\
		      \ \ \ \ \                 $z$ = WeightedCenterLayer($[z_t^1,z_t^2]$, $c_i$ ,$w_i$) \\
		    \ \ \ \ \ $s_z$ = SoftMaxLayer($z$)     \\
		     \ \ \ \ \ Loss = CrossEntropyLoss($s_z$) + 0.3 $\times$ CenterLoss($z$) \\
		
		     \ \ \ \ \ Loss.backward()\\
		     \ \ \ \ \ Optimizer.step()
		
		\caption{Pseudocode for Our Proposed Algorithm}
		\label{alg:alg1}
	\end{algorithm}

	%% +++++++++++++++++++++++++++++++++++++++++++++++++++++++
	%%
	%% 3.3 SoftMax with Weighted Center Loss
	%%
	%% +++++++++++++++++++++++++++++++++++++++++++++++++++++++
	\subsection{Classification}
	\label{ssec:softmax_with_weighted_center_loss}
	
	A softmax loss function alone cannot recognize intra-class variance (i.e., differences of features within a class) and inter-class variance (i.e., differences of features across classes).
To overcome this problem, we first use a center loss function proposed by Wen et al. \cite{wen2016discriminative} to make the distance of the features belonging to different classes be further away while the distance of the features belonging to the same class be closer.
	\begin{equation}
		L_{s} = -\frac{1}{N}\sum ^N_{i=1}\sum ^T_{t=1}log\left[\frac{exp\left(F\left(x_i;\theta_t\right)\right)\cdot y_t^i}{\sum ^T_{j=1}exp\left(F\left(x_i;\theta_t\right)\right)}\right]
	\end{equation}
	where $x_i$ represents the $i$th sample in the dataset of the size $N$. $y_t\in\{0, 1\}^t $ indicates the one-hot encoding applied to the input based on the labels. $T$ indicates the number of tasks during training (e.g., either in the entire dataset or in the mini-batch).
	
	Additionally, we also use the second term of center loss to create clusters for different ransomware families. The distance across the training samples belonging to the same class is shortened around each cluster while the distance across the training samples of different classes is separated. However, it also implies that in the course of learning a center cluster and the feature $x_i$, they may have a similar direction in the feature embedding during the gradually optimizing step. To mitigate the redundancy in the learning procedure, we assign the different weights on each class cluster, which is formally defined as follows.
	\begin{equation}
		L_{c}=\frac{1}{2N}\sum_{N}^{i=1}\left\| x_i-w_{y_t}c_{y_t} \right\|_{2}^{2}
		\label{eq:center}
	\end{equation}
	where $c_{y_t}$ denotes the center of a class $i$ while $x_i$ denotes the features of the euclidean distance and $w_i$ indicates the class weights \cite{king2001logistic} on each center cluster.

The objective function stands for the squared Euclidean distance. Observed from Equation (\ref{eq:center}), the center loss with different weight parameters encourages instances of the same class to be closer to a learned class center by different gradient descent directions of each class.

	The classification probability is calculated by the use of two terms of center loss and the softmax loss function, as shown below.
	\begin{equation}
	    \begin{split}
    		\mathit{Classification} =  &-\frac{1}{N}\sum ^N_{i=1}\sum ^T_{t=1}log\left[\frac{exp\left(F\left(x_i;\theta_t\right)\right)\cdot y_t^i}{\sum ^T_{j=1}exp\left(F\left(x_i;\theta_t\right)\right)}\right] \\ &+ \alpha \frac{1}{2N}\sum_{N}^{i=1}\left\| x_i-w_{y_t} c_{y_t} \right\|_{2}^{2}
		\end{split}
	\end{equation}
	where $\alpha$ is the hyperparameter to balance the two terms.

	%%%%%%%%%%%%%%%%%%%%%%%%%%%%%%%%%%%%%%%%%%%%%%%%%%%%%%%%%%%%%%%
	%%
	%%  4. Experimental Setup and Evaluation Metrics
	%%
	%%%%%%%%%%%%%%%%%%%%%%%%%%%%%%%%%%%%%%%%%%%%%%%%%%%%%%%%%%%%%%%
	\section{Experiment Setup and Evaluation Metrics}
	\label{sec:experimental_setup_and_evaluation_metrics}
	
	This section describes the dataset, the experiment setup, and the performance metrics used to evaluate the proposed approach.

	%% +++++++++++++++++++++++++++++++++++++++++++++++++++++++
	%%
	%% 4.1 Dataset Description
	%%
	%% +++++++++++++++++++++++++++++++++++++++++++++++++++++++
	\subsection{Dataset Description}
	\label{ssec:dataset_description}
	
	In order to assess the effectiveness of the proposed method, the following benchmark dataset is used in this study.
	
	The binaries of this dataset’s instances were obtained from ViruseShare. The dataset comprises 11 families/classes of ransomware, each of which consists of a varying number of instances as listed in Table \ref{tbl:ransomware_dataset}. Clearly, the dataset is highly imbalanced, which reflects reality, as some classes, e.g., Petya and Dalexis, are largely outnumbered by the other classes, e.g., Zerber, as shown in the third column of Table  \ref{tbl:ransomware_dataset}.
	
	%As a way of evading static-based techniques of malware analysis, many malware authors utilise packing techniques in order to hide informative static indicators, e.g., strings, which can be used to detect such malware. Packing software utilise compression techniques in order to reduce the size of the executable binary. As a result of this compression process, many useful strings will be encoded differently, which can be accessed again once the underlying executable is unpacked, i.e., decompressed.
	
	%Identifying packed malware can be an easy task if well-known packers, such as Ultimate Packer for eXecutables (UPX), are used mainly due to the fact that most of these packing software will add some identifiable strings, e.g., UPX0, that need to be relied on to unpack the executable. Relying on such strings can largely help malware analysts to recognise that the executable is packed and which packing software can be used to unpack it. The fourth column in Table \ref{tbl:ransomware_dataset} shows the percentage of packed instances in each group using only those well-known packers. However, many malware authors nowadays may develop their own packing software to reduce the possibility of their malware being identified as being packed and/or unpacked. This issue, however, can be resolved by utilising dynamic analysis techniques as the malware needs to unpack itself to be executed on the system. Dynamic analysis techniques of malware are not within the scope of this study.
	
	\begin{table}[th!]
		\caption{Details of the ransomware dataset.}
		\label{tbl:ransomware_dataset}
		\centering
		{\footnotesize
			\begin{tabular*}{0.98\textwidth}{@{\extracolsep{\fill}}lcc}
				\toprule
				
				Family & Instances & Ratio (\%) %& Packed  (\%)
				\\
				\midrule
				Bitman & 99 & 9.45 %& 19.2
				\\
				Cerber & 91 & 8.68 %& 30.8
				\\
				Dalexis & 9 & 0.86 %& 11.1
				\\
				Gandcrab & 100 & 9.54 %& 1.00
				\\
				Locky & 96 & 9.16 %& 35.4
				\\
				Petya & 6 & 0.57 %& 0.00
				\\
				Teslacrypt & 91 & 8.68 %& 28.6
				\\
				Upatre & 18 & 1.72 %& 5.60
				\\
				Virlock & 162 & 15.46 %& 8.00
				\\
				Wannacry & 178 & 16.98 %& 49.4
				\\
				Zerber & 198 & 18.89 %& 10.1
				\\
				
				\bottomrule			
			\end{tabular*}
		}
	\end{table}

    %% +++++++++++++++++++++++++++++++++++++++++++++++++++++++
	%%
	%% 4.2 Experimental Setup
	%%
	%% +++++++++++++++++++++++++++++++++++++++++++++++++++++++
    \subsection{Experimental Setup}
    \label{ssec:experimental_setup}

    This study was carried out using a 3.6 GHz 8-core Intel Core i7 processor with 32 GB memory on Windows 10 operating system. The proposed approach is developed using Python programming language with several statistical and visualization packages such as Sckit-learn, Numpy, Pandas, Pytorch, and Matplotlib. Table~\ref{table:Mat} summarizes the system configuration.
    %This work comprised of six steps. Firstly, exploratory data analysis was carried out on training dataset to remove or replace any missing values in dataset. Secondly, data is preprocessed to convert all the categorical variables into one-hot encoded variables. Thirdly, using statistical filtering less useful features and samples are filtered out from  dataset. In the fourth step,  conditional variational autoencoder model is developed to learn latent structure of intrusion samples.  In the fifth step, we generated samples less in the dataset to balance the dataset. Finally, latent embedding is classified into instrusion samples using stacked auto encoder.

    \begin{table}[th!]
    	\caption{Implementation environment specification.}
    	\label{table:Mat}
    	\centering
    	{\footnotesize
    		\begin{tabular*}{0.98\textwidth}{@{\extracolsep{\fill}}lp{8cm}}
    			\toprule
    			\textbf{Unit}   & \textbf{Description}\\
    			\midrule
    			Processor   & 3.6 GHz 8-core Inter Core i7 \\
    			RAM  &  32 GB      \\
    			GPU  & GeForce GTX 1080 Ti       \\
    			Operating System  & Windows 10  \\	
    			Packages   &  Pytorch, Sckit-Learn, Numpy, Pandas, Pyriemannian and Matplotlib    \\
    			\bottomrule
    		\end{tabular*}
    	}
    \end{table}

    %\label{sec:guidelines}

    %% +++++++++++++++++++++++++++++++++++++++++++++++++++++++
	%%
	%% 4.3 Evaluation Metrics
	%%
	%% +++++++++++++++++++++++++++++++++++++++++++++++++++++++
    \subsection{Evaluation Metrics}
    \label{ssec:evaluation_metrics}

    The proposed method is compared and evaluated using Accuracy, Precision, Recall, F1-score, and Area under the receiver operating characteristics (ROC) curve. In this work, we have used the macro and micro average of Recall, Precision, and F1-score for multi-class classification. All the above metrics can be obtained using the confusion matrix (CM) which is described in Table ~\ref{tab:conmat}.

    \begin{table}[th!]
		\caption{Illustration of confusion matrix.}
		\label{tab:conmat}
		\centering
		{\footnotesize
			\begin{tabular}{l|l|c|c|}
				\multicolumn{2}{c}{}&\multicolumn{2}{c}{Predicted} \\
				\cline{3-4}
				\multicolumn{2}{c|}{} & Class$_{\mathit{pos}}$ & Class$_{\mathit{neg}}$ \\
				\cline{2-4}
				\multirow{2}{*}{Actual} & Class$_{\mathit{pos}}$ & True Positive (\textit{TP}) & False Positive (\textit{FP}) \\
				\cline{2-4}
				& Class$_{\mathit{neg}}$ & False Negative (\textit{FN}) & True Negative (\textit{TN}) \\
				\cline{2-4}
			\end{tabular}
			%			\begin{tabular*}{0.98\textwidth}{@{\extracolsep{\fill}}cccc}
				%				\toprule
				%				\multicolumn{1}{l}{}    &         & \multicolumn{2}{c}{Predicted}        \\\cmidrule(l){2-4}  %\midrule
				%				\multicolumn{1}{l}{}    &         & Class$_{pos}$ & Class$_{neg}$ \\ \midrule
				%				\multirow{2}{*}{Actual} & Class$_{pos}$ & True Positive $(TP)$     & False Positive $(FP)$      \\
				%				& Class$_{neg}$ & False Negative $(FN)$     & True Negative $(TN)$      \\
				%				\bottomrule
				%			\end{tabular*}
		}
	\end{table}
    In Table~\ref{tab:conmat}, True positive (TP) means the amount of class$_{pos}$ data predicted actual belong to class$_{pos}$, True negative (TN) is amount of class$_{neg}$ data predicted is actually class$_{neg}$, False positive (FP) indicates data predicted class$_{pos}$ is actual belong to class$_{neg}$ and False negative (TN) is data predicted as class$_{neg}$ but actually belong to class$_{pos}$. Based on the aforementioned terms, the evaluation metrics are calculated as follows.
    %Accuracy (ACC) measures the total number of data samples that are correctly classified as shown in Equation ~\ref{eq:ACC}.
    For a balanced test dataset, higher accuracy indicates the model is well learned, but for unbalanced test dataset scenarios relying on accuracy can give the wrong illusion about the model's performance. Thus for the unbalanced datasets, recall and F1-score metrics give a more intuitive explanation of the model's performance.

   %\begin{equation}\label{eq:ACC}
   %	ACC = \frac{TP+TN}{TP + TN + FP + FN}
   % \end{equation}
   % }
%    As the dataset is highly-imbalanced, the balanced accuracy ($\mathit{Accuracy}$) is used instead of the typical accuracy. The balanced accuracy formally defined for two class problem as:
    %
    %\begin{equation}
    %	\mathit{Accuracy}=\frac{1}{n}\sum_{i=1}^n\frac{\mathit{correct}_i}{\mathit{total}_i},
  %  \end{equation}
  %    \begin{equation}\label{eq:ACC}
   % 	Accuracy = \frac{\frac{TP}{Class_{pos row}}+\frac{TN}{Class_{neg row}}}{2}
  %  \end{equation}
    %where $n$ is the number of classes, and $\mathit{correct}_i$ and $\mathit{total}_i$ is, respectively, the number of correctly classified instances and total number of instances of the $i$th class.
    %Balanced Accuracy is essentially an average of recalls. First we evaluate recall for each class and the we average the values in order to obtain balanced accuracy. Above definition can be extended for unbalanced multi-class problem.
    Recall (also known as true positive rate) estimates the ratio of the correctly predicted samples of the class to the overall number of instances of the same class. It can be computed using Equation ~\ref{eq:TPR}. A higher Recall value indicates the good performance of a model.
    \begin{equation}\label{eq:TPR}
    	\mathit{Recall} = \frac{\mathit{TP}}{\mathit{TP} + \mathit{FN}}
    \end{equation}

    Precision measures the quality of the correct predictions. Mathematically, it is the ratio of correctly predicted samples to the number of all the predicted samples for that particular class as shown in Equation~\ref{eq:PPV}. Precision is usually paired with Recall to evaluate the performance of a model. Sometimes pairs can appear contradictory thus comprehensive measure F1-score is considered for unbalanced test data-sets.
    \begin{equation}\label{eq:PPV}
    	\mathit{Precision} = \frac{\mathit{TP}}{\mathit{TP} + \mathit{FP}}
    \end{equation}

    F1-score computes the trade-off between precision and recall. Mathematically, it is the harmonic mean of precision and recall, which is computed as
    \begin{equation}\label{eq:F-measure}
    	F1-score = 2\times\left(\frac{\mathit{Precision}\times\mathit{Recall}}{\mathit{Precision} + \mathit{Recall}}\right)
    \end{equation}

    The area under the curve (AUC) computes the area under the receiver operating characteristics (ROC) curve which is plotted based on the trade-off between the true positive rate on the y-axis and the false positive rate on the x-axis across different thresholds \cite{Goodfellow-et-al-2016}. Mathematically, AUC is computed as
    \begin{equation}\label{eq:auc}
    	\mathit{AUC}_{\mathit{ROC}}=\int_{0}^{1} \frac{\mathit{TP}}{\mathit{TP}+\mathit{FN}}d\frac{\mathit{FP}}{\mathit{TN}+\mathit{FP}}
    \end{equation}

    In the case of multi-class and unbalanced test data classification, the performance of models is usually evaluated using weighted and micro-averaging of recall, precision, and F1-score. We have employed weighted averaging which
    % as shown in the set of equations ~\ref{eq:macro} and ~\ref{eq:micro} respectively.
    %\begin{equation}
    	%\begin{aligned}
    		%R_{macro} &= \frac{1}{m}\sum_{i=i}^{m}\frac{TP_i}{TP_i+FN_i}\\
    		%\\
    		%P_{macro} &= \frac{1}{m}\sum_{i=i}^{m}\frac{TP_i}{TP_i+FP_i} \\
    		%\\
    		%F_{macro} &= 2\times\frac{R_{macro}\times P_{macro}}{R_{macro}+ P_{macro}}
    		%\end{aligned}
	%\label{eq:macro}
	%\end{equation}
  %Macro-averaging
  in simple terms is the  weighted mean of recall, precision, and F1-scores with weights equal to class probability while summing up the individual recall, precision, or F1-scores in each class respectively,  as shown in Equations~(\ref{eq:r_micro})--(\ref{eq:f_micro}).

    \begin{align}
    		R_{\mathit{weighted}} &= \sum^{m}_{i=1}\frac{n_i}{\sum_{j=1}^mn_j}R_i\label{eq:r_micro}\\
    		%\\
    		P_{\mathit{weighted}} &= \sum^{m}_{i=1}\frac{n_i}{\sum_{j=1}^mn_j}P_i \label{eq:p_micro}\\
    		F_{\mathit{weighted}} &= \sum^{m}_{i=1}\frac{n_i}{\sum_{j=1}^mn_j}F_i \label{eq:f_micro}
	\end{align}

    %%%%%%%%%%%%%%%%%%%%%%%%%%%%%%%%%%%%%%%%%%%%%%%%%%%%%%%%%%%%%%%
	%%
	%%  5. Results and Discussion
	%%
	%%%%%%%%%%%%%%%%%%%%%%%%%%%%%%%%%%%%%%%%%%%%%%%%%%%%%%%%%%%%%%%

    \section{Experimental Results}
    \label{sec:results_and_discussion}

We evaluated our proposed model %\textcolor{blue}{, which is included of 41 million parameters and numerical computing performance is 3.8G flops, }
on ransomware dataset summarized in Table \ref{tbl:ransomware_dataset}. We benchmarked our model with both the lightweight deep learning models (e.g., deep neural network (DNN) and recurrent neural network (RNN))  and the heavyweight deep learning models (e.g., InceptionV3, Resnet50, and VGG16) used widely in malware classification tasks that are trained with the limited number of samples.

%We conducted the experiments on 755 samples of ransomware dataset and compared with the baseline methods of
 %   \begin{table}[ht!]
%		\caption{performance of different models}
%		\label{tab:results}
%		\centering
%		{\footnotesize
%			\begin{tabular*}{0.98\textwidth}{@{\%extracolsep{\fill}}lllllll}
		%		\toprule
				
%				Ref.    & Method &  Accuracy & Recall & Precision & F-score & Auc-Roc \\
    %		\cite{yue2017imbalanced}      & %VGG16   &  74.6 &   74.6 &   77.1   & 75.2 &  94.2\\
    %		\cite{chen2018deep}   & Inception V3 %    & 74.8 & 74.8 & 78.6  &76.3& 96.1\\
    %		\cite{8763852}      & Xception & 578.5 & 73.4 & 77.8 &  76.5 &  96.3\\
  %  		    & Our model  & 88.7 & 88.7 &  %85.3 & 86.2 & 98.2 \\
    		
%				\bottomrule
%			\end{tabular*}
	%	}
%	\end{table}
	
	%%%% execution time
In order to have an unbiased comparison, all the models are trained under bootstrap sampling with 30 repetitions using   80\% ransomware data and the remaining 20\% is used for evaluating their performances.

Table \ref{tab:results} shows a performance comparison between our proposed approach and other state-of-the-art approaches used as a baseline to benchmark our results.
All models in Table \ref{tab:results} were trained with entropy features generated from binary executables. As it can be seen in Table \ref{tab:results}, our approach exceeds DNN by14.5\%, RNN by13.3\%, VGG16 by 14.1\%, Inception V3 by 13.9\%, and Xception by 15.3\% in terms of weighted average recall metric respectively. In all the metrics our approach outperformed in comparison to other baseline models.
	   \begin{table}[th!]
		\caption{performance of different models}
		\label{tab:results}
		\centering
		{\footnotesize
			\begin{tabular*}{0.98\textwidth}{@{\extracolsep{\fill}}llllll}
				\toprule
				
				Ref.    & Method &   Recall & Precision & F1-score & Auc-Roc
				\\
    		\midrule
    			\cite{basnet2021ransomware}   & DNN &  74.2 & 76.6  &75.3& 95.8  \\
    		\cite{basnet2021ransomware}      & RNN &  75.4 & 77.8 &  76.1 &  96.1\\
    		\cite{Yue08042}      & VGG16   &     74.6 &   77.1   & 75.2 & 94.2 \\
    		\cite{lichen}   & Inception V3     &  74.8 & 78.6  &76.3& 96.1 \\
    		\cite{8763852} & Xception &  73.4 & 77.8 &  76.5 &  96.3\\
    	     & Our model (entropy features)  & 88.7 &  85.3 & 86.2 & 98.2\\
    		    & Our model (gray scale images)  & 82.9 &  80.3 & 81.8 & 94.5 \\
    		
				\bottomrule
			\end{tabular*}
		}
	\end{table}

\begin{figure}[th!]
    	\centering
    	\includegraphics[scale=0.35]{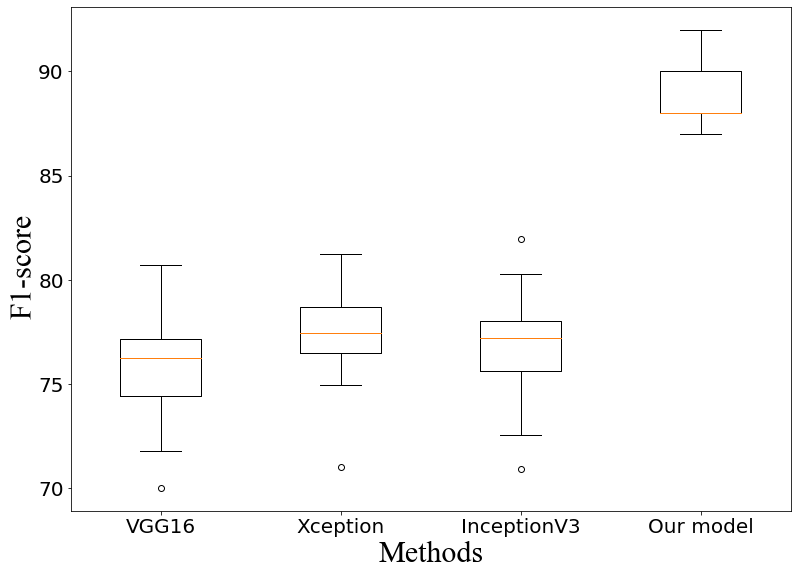}
    	\caption{Boxplot showing performance variation of different methods under bootstrap sampling with 80\% training and 20\% test data}
    	\label{fig:methods_boxplot}
    \end{figure}

Figure \ref{fig:methods_boxplot} shows performance variations in terms of F1-score metric for different models under 30 repetitions. We can see at 80\% of training data size performance variation (with 20\% test set) is very minimal. We were able to achieve a mean F1-score of more than 88\% with the lowest 86\% and highest around 94\%. Other models in some repetitions are able to achieve above 80\% weighted F1-score but the average is around 75\%.

\begin{figure}[th!]
    	\centering
    	\includegraphics[scale=0.55]{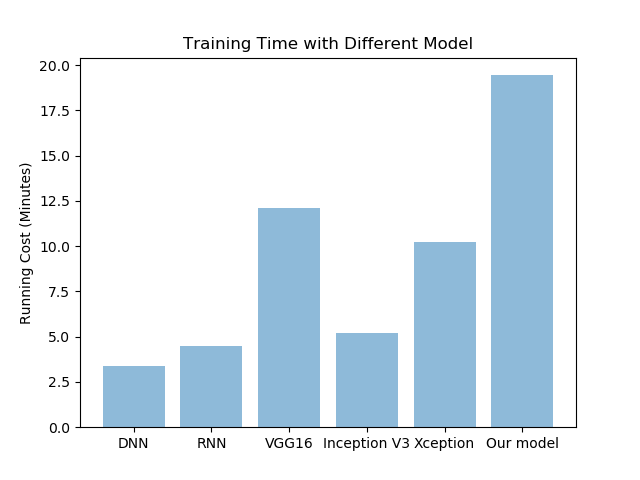}
    	\caption{The training time compared with DNN, RNN, VGG16(Fine-tune), Inception V3(Fine-tune), Xception(Fine-tune)}
    	\label{fig:train_different}
    \end{figure}

\textcolor{blue}{
We also compared computation cost in terms of training time for each model based on entropy features in Table \ref{tab:results}. Figure \ref{fig:train_different} shows the training time of our proposed model compared to all the other similar models. The training time for our proposed model is higher because our model uses more trainable parameters.
}

Table \ref{tab:parameter1} lists hyper-parameter setting for baseline and our proposed approach. For instance, all the models were trained with $50$ epochs.
	\begin{table}[th!]
		\caption{Hyper-Parameter setting for different model}
		\label{tab:parameter1}
		\centering
		{\footnotesize
			\begin{tabular*}{0.98\textwidth}{@{\extracolsep{\fill}}lll}
				\toprule
    		Training Parameters& Other models  & Our model  \\
    		\midrule
    		Epoch Number &  50 &  50    \\
    		Batch Size  &  32 &24   \\
    		Optimizer &  Adam& Adam   \\
    		Learning rate &  1e-4& 1e-4   \\
    		Train/Test&  80\%/20\% & 80\%/20\% \\
    		Re-scaling&  1/255 & 1/255   \\
    		Augmentation&  for 3 classes  &  for 3 classes  \\
    	
    		Loss function&  Crossentropy & Crossentropy  \\
    	& & Center loss  \\
				\bottomrule

			\end{tabular*}
		}
	\end{table}

\begin{figure}[th!]
    	\centering
    	\includegraphics[scale=0.4]{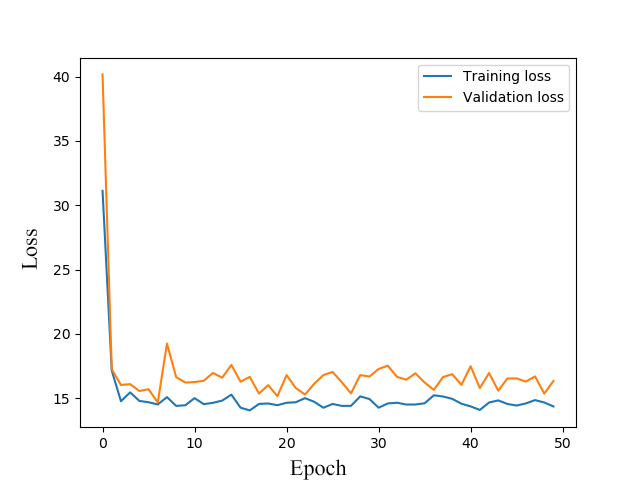}
    	\caption{Loss during Training on Epochs}
    	\label{fig:loss}
    \end{figure}

As Figure  \ref{fig:loss} illustrates, both train and validation loss stabilizes during 50 epochs confirming the training is done by this stage.
As we can see by classifying the input samples into 11 ransomware classes, our model has achieved an F1-score of more than 86\%. As observed from Table \ref{tbl:ransomware_dataset}, the sample number of dalexis, petya, and uptatre is less compared with other classes, which results in the imbalanced problem in the training stage. To solve this issue, we employ the data augmentation technique \cite{zhu2020multi}  for these 3 classes and achieved better results. % as shown in hyper-parameter settings in  Table \ref{tab:parameter1}.

\begin{figure}[th!]
    	\centering
    	\includegraphics[scale=0.5]{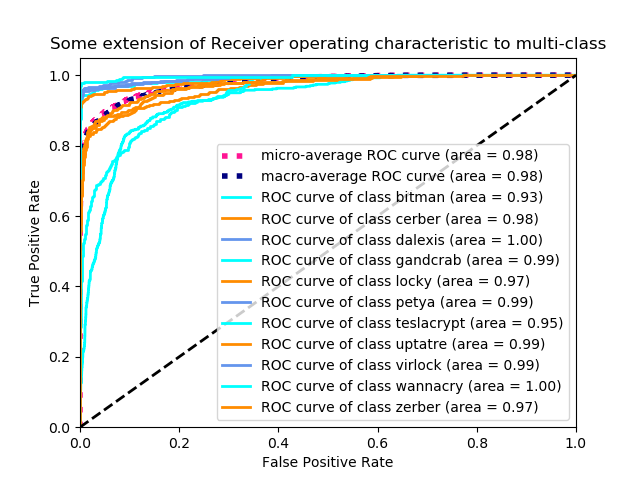}
    	\caption{Area under curve performance our proposed approach}
    	\label{fig:Roc}
    \end{figure}

\begin{figure}[th!]
    	\centering
    	\includegraphics[scale=0.5]{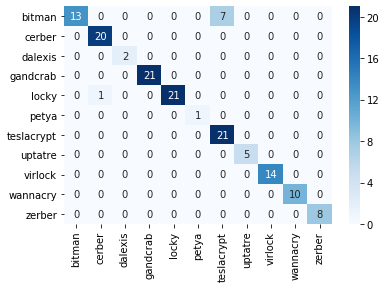}
    	\caption{Confusion Matrix}
    	\label{fig:confusion_matrix}
\end{figure}

\begin{figure}[th!]
    	\centering
    	\includegraphics[scale=0.30]{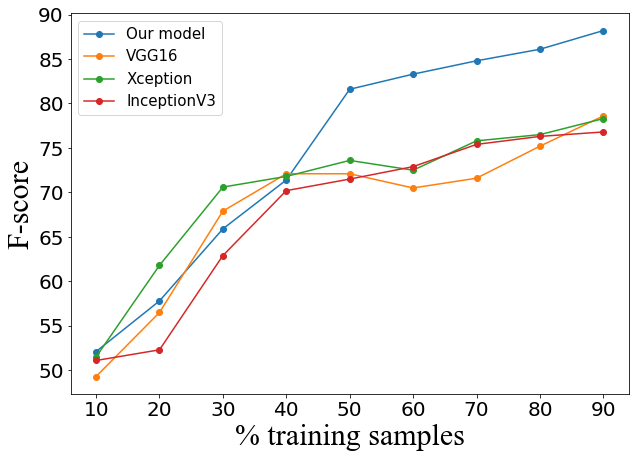}
    	\caption{Performance with the different ratio of \% training samples}
    	\label{fig:ratio_training}
    \end{figure}

Figure \ref{fig:Roc} shows the area under the curve for different ransomware classes in receiver operating characteristics (ROC). The micro and macro average of all the classes is around 0.98 which indicates our model has good separability between different classes.\\
Furthermore, we can see different clusters formed under our model after training.
	\begin{figure}[th!]
    	\centering
    	\subfloat[Epoch 1] {\includegraphics[width=0.4\textwidth]{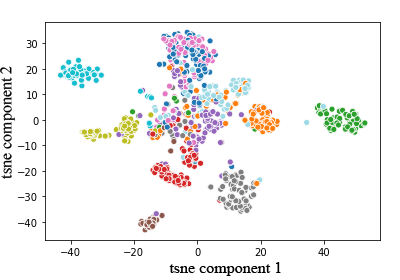}} \hfil
    	\subfloat[Epoch 50] {\includegraphics[width=0.5\textwidth]{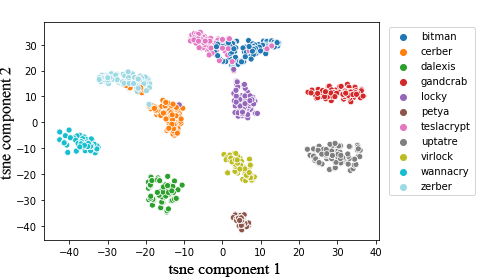}} \hfil
    	\caption{T-SNE visualization of our data before and after training}
    	\label{fig:tsne}
    \end{figure}	
Figure \ref{fig:tsne} shows different clusters of ransomware classes in our data that are visualized using the t-sne dimensionality reduction technique. Firstly, the t-sne plot in Figure \ref{fig:tsne} indicates that entropy is a very informative feature as we can see different clusters of ransomware attacks at the beginning of our model's training. Secondly, some of the ransomware attack classes are easily distinguishable even at the first epoch of training such as locky,  wannacry, uptatre, zerber and virlock, etc. At the end of the training, we can see our model has learned feature embedding to categorize ransomware samples based on their similarity with the cluster.  This can also be seen in our proposed approach's results shown by the confusion matrix for one run in bootstrap sampling in Figure \ref{fig:confusion_matrix}.\\
As we can see in Figure \ref{fig:confusion_matrix}, our model is able to distinguish all the classes of ransomware except Bitman and Teslacrypt. This is because these two classes have very similar signatures that can be seen in the t-sne plot after 50 epochs in Figure \ref{fig:tsne}. Both classes clusters are very close to each other with some samples very difficult to differentiate, that is why our model performed poorly for these two classes.
%Our model was able to distinguish different ransomware attack classes with almost zero false positives whereas there is a trade-off between Bitman and teslacrypt ransomware classes. %The Figure \ref{fig:confusion_matrix} confusion matrix summarizes the ransomware classification performance of a classifier with respect to some test data.
We further conduct another experiment to explore the performance of our model with different ratios of training size, as shown in Figure \ref{fig:ratio_training}.

%\begin{figure}[ht!]
%    	\centering
%    	\includegraphics[scale=0.30]{accuracy_booster.png}
%    	\caption{Performance with the different ratio of \% training samples}
%    	\label{fig:ratio_training}
 %   \end{figure}
We systematically reduce the training data size from 90\% to 10\%. This experiment was repeated 30 times using bootstrapping technique (sampling with replacement) and the mean F1-score on test data based on training data size is reported using Figure \ref{fig:ratio_training}. As we can see, our model is able to achieve a weighted F1-score greater than 80\% at just 50\% training size whereas other state-of-the-art methods required a very large amount of data (approx. 90\% of data for training) to reach more than 70\%  weighted F1-score.   Finally, from all these experiments we can confirm our model performs well under a small amount of training data.
%Moreover, our training method is not flexible compared to online learning methods \cite{sahoo2017online}. Nowadays, an ever-increasing number of new ransomware variants become a threat to cybersecurity and web analytics. Online learning usually focuses on learning a single instance and is adaptive because once the data is consumed, it is no longer needed. In our scenario, our proposed model still trains using the method that consumes a number of examples compared to most online learning methods.
 %\begin{figure}[ht!]
%	\centering
  %	\includegraphics[scale=0.3]{ratio_training.png}
  %	\caption{Boxplot showing performance variation of our model under different settings}
 %  	\label{fig:boxplot}
 %   \end{figure}

Although the proposed model performs well overall, it has several limitations. First of all, our model is still unable to classify the Bitman ransomware family or the Teslacrypt ransomware family, as shown in Table \ref{tab:results}. The reason for this can be seen in Figure \ref{fig:tsne} where after training, the clusters of Bitman and Teslacrypt families still overlap. Without a clear cluster around these two families, our model tends to produce mi-classification results especially across the malware instances belonging to these two families. Second of all, our proposed model has not been tested for zero-day attacks but only provides results for the classification of known attacks that are included in the training samples. Finally, our model takes a longer training time compared to other methods due to the higher number of parameters that need to be trained.

    %%%%%%%%%%%%%%%%%%%%%%%%%%%%%%%%%%%%%%%%%%%%%%%%%%%%%%%%%%%%%%%
	%%
	%%  6. Conclusions
	%%
	%%%%%%%%%%%%%%%%%%%%%%%%%%%%%%%%%%%%%%%%%%%%%%%%%%%%%%%%%%%%%%%
    \section{Conclusions}
    \label{sec:conclusions}

    We proposed a few-shot learning method based on the Siamese Neural Network model which can rapidly detect and classify different ransomware attack classes even if only a few ransomware samples are available for each class. In our proposed approach, we use entropy features in addition to image features to capture more accurately the unique characteristics associated with each malware signature represented in the feature embedding space of the model. A pre-trained network such as VGG-16 is utilized to generate more accurate weights of different ransomware samples in a meta-learning fashion to avoid the bias typically associated with a model trained with a limited number of training samples. The experimental results tested on the ransomware samples containing 11 different ransomware classes show a very high classification with a weighted F1-score exceeding 86\%.

    We plan to extend our work to other application domains to evaluate the generalizability and practicability of our proposed solution. These include but are not limited to the rapid classification of other types of malware \cite{mcintosh2018large, wei2021aemlp, xu2021improving, mcintosh2019inadequacy}. We also plan to extend our work to a practical market-scale triage solution to screen different malware rapidly.

%\vspace{-0.25cm}
%\bibliographystyle{IEEEtran}
%\bibliography{IEEEabrv,SystemModelQoE}

    %%%%%%%%%%%%%%%%%%%%%%%%%%%%%%%%%%%%%%%%%%%%%%%%%%%%%%%%%%%%%%%
	%%
	%%  Acknowledgment
	%%
	%%%%%%%%%%%%%%%%%%%%%%%%%%%%%%%%%%%%%%%%%%%%%%%%%%%%%%%%%%%%%%%
    \section*{Acknowledgment}
    This research is supported by the Cyber Security Research Programme—Artificial Intelligence for Automating Response to Threats from the Ministry of Business, Innovation, and Employment (MBIE) of New Zealand as a part of the Catalyst Strategy Funds under the
    grant number MAUX1912.

\bibliography{Bibliography}

\end{document}